\documentclass[pra,aps,twocolumn,preprintnumbers,superscriptaddress]{revtex4}
\usepackage{latexsym,epsfig,amssymb,amsfonts,amsmath,graphicx,bbm}

\newcommand{\half}{\mbox{$\textstyle \frac{1}{2}$}}
\newcommand{\ket}[1]{\left | \, #1 \right \rangle}
\newcommand{\kets}[1]{| \, #1 \rangle}

\newcommand{\av}[1]{\langle #1\rangle}
\newcommand{\tr}[1]{\mbox{tr} \, #1 }

\newcommand{\eqr}[1]{Eq.~\ref{#1}}
\newcommand{\secr}[1]{Sec.~\ref{#1}}
\newcommand{\fir}[1]{Fig.~\ref{#1}}

\allowdisplaybreaks

\begin{document}

\title{Re-entrance and entanglement in the one-dimensional Bose-Hubbard model}
\author{M. \surname{Pino}}
\affiliation{Departamento de F\'{\i}sica - CIOyN, Universidad de Murcia, Murcia 30071, Spain} 
\author{J. \surname{Prior}}
\affiliation{Departamento de F\'{\i}sica Aplicada, Universidad Polit\'ecnica de Cartagena, Cartagena 30202, Spain}
\author{A.~M. Somoza}
\affiliation{Departamento de F\'{\i}sica - CIOyN, Universidad de Murcia, Murcia 30071, Spain}
\author{D. \surname{Jaksch}}
\affiliation{Clarendon Laboratory, University of Oxford, Parks Road, Oxford OX1 3PU, U.K.}
\affiliation{Centre for Quantum Technologies, National University of Singapore, 2 Science Drive 3, Singapore 117542}
\author{S.~R. \surname{Clark}}
\affiliation{Centre for Quantum Technologies, National University of Singapore, 2 Science Drive 3, Singapore 117542}
\affiliation{Clarendon Laboratory, University of Oxford, Parks Road, Oxford OX1 3PU, U.K.}

\date{\today}

\begin{abstract}
Re-entrance is a novel feature where the phase boundaries of a system exhibit a succession of transitions between two phases A and B, like A-B-A-B, when just one parameter is varied monotonically. This type of re-entrance is displayed by the 1D Bose Hubbard model between its Mott insulator (MI) and superfluid phase as the hopping amplitude is increased from zero. Here we analyse this counter-intuitive phenomenon directly in the thermodynamic limit by utilizing the infinite time-evolving block decimation algorithm to variationally minimize an infinite matrix product state (MPS) parameterized by a matrix size $\chi$. Exploiting the direct restriction on the half-chain entanglement imposed by fixing $\chi$, we determined that re-entrance in the MI lobes only emerges in this approximate when $\chi \geq 8$. This entanglement threshold is found to be coincident with the ability an infinite MPS to be simultaneously particle-number symmetric and capture the kinetic energy carried by particle-hole excitations above the MI. Focussing on the tip of the MI lobe we then applied, for the first time, a general finite-entanglement scaling analysis of the infinite order Kosterlitz-Thouless critical point located there. By analysing $\chi$'s up to a very moderate $\chi = 70$ we obtained an estimate of the KT transition as $t_{\rm KT}=0.30\pm 0.01$, demonstrating the how a finite-entanglement approach can provide not only qualitative insight but also quantitatively accurate predictions.
\end{abstract}

\maketitle

\section{Introduction}
Strong correlations in many-body quantum systems are central to the appearance of numerous remarkable phenomenon such as the fractional quantum effect~\cite{Tsui82} and high-temperature superconductivity~\cite{Bednorz86}. As such the study of model Hamiltonians composed of spins, fermions and bosons has played an crucial role in unravelling the fundamental mechanisms underlying them~\cite{Sachdev01,Leggett06}. Recently the relevance of these types of model systems has been dramatically elevated by numerous seminal experiments with cold atoms in optical lattices~\cite{Greiner02,Stoferle04,Spielman07,Gemelke09,Bakr10,Haller10,Trotzky10}. In the simplest instances these experiments provided a clean and highly controllable quantum degenerate atomic system whose microscopic interactions are quantitatively described by the Bose-Hubbard model (BHM)~\cite{Jaksch98,Bloch08}. 

The BHM is a minimal many-body Hamiltonian that contains the key physics of strongly interacting soft-core bosons on a lattice. It embodies the competition between the kinetic and repulsive on-site interaction energies giving rise to a quantum phase transition. For small interactions the bosons are completely delocalized leading to a superfluid (SF) phase, while for sufficiently large interactions, and a commensurate density, the bosons become localized and enter the Mott insulator (MI) phase. By increasing the depth of the optical lattice potential this archetypal SF to MI transition has been experimentally observed in cold-atom systems with one dimensional (1D)~\cite{Stoferle04,Haller10}, 2D~\cite{Spielman07,Gemelke09,Bakr10}, and 3D~\cite{Greiner02,Trotzky10} lattices.

The essential qualitative features of the BHM phase diagram, such as the existence of MI lobes, depicted in \fir{fig:mf_phase_diagram}, were worked out some time ago by Fisher {\em et al}.~\cite{Fisher89}. Nonetheless the study of the BHM's SF-MI transition continues to attract much attention~\cite{Anderson11}, with a large body of work~\cite{Batrouni90,Batrouni92,Kashurnikov96a,Singh92,Freericks96,Elstner99b,Kuhner98,Kuhner00} attempting to enhance the quantitative understanding of its structure. The focus of the work described here is on the BHM in 1D, which in many respects displays rather peculiar physics. For example, in 1D the SF phase is not a Bose condensate of the single particle state with the lowest kinetic energy, but is instead characterised by an algebraic diverging momentum distribution~\cite{Giamarchi04}. Moreover, when crossing the tip of a MI lobe in 1D the energy gap closes exponentially slowly reflecting the Kosterlitz-Thouless (KT)~\cite{Fisher89,Kosterlitz73} nature of the transition there. Related to this, and exclusively in 1D, the shape of the MI lobes also displays a novel and unexpected feature known as re-entrance~\cite{Elstner99b,Kuhner98,Kuhner00,Kashurnikov96a}, as shown most clearly in \fir{fig:phase_diagrams}(d). 

In general, a system exhibits re-entrance when a succession of transitions between two phases A and B, such as A-B-A-B, can occur by monotonically increasing just one parameter. Such a sequence is often counter-intuitive. For example in the context of classical thermal phase transitions it is natural for the varying parameter to be temperature. It is then expected that the low temperature phase A will be ordered, while increasing the temperature will drive the system to a disordered phase B. However, the appearance of re-entrance means that increasing the temperature can in fact unexpectedly stabilize the ordered phase A again. Precisely this sequence of phase transitions has been observed in liquid crystals between the A = Smectic (ordered) phase and B = Nematic (disorded) phase with increasing temperature~\cite{Cladis75,Chandrasekhar92}. Reentrance has also been predicted to occur in classical frustrated spin systems through a mechanism of ``order by disorder"~\cite{Saslow86}.

At some constant chemical potential the BHM in 1D displays a similar re-entrant sequence of zero-temperature quantum phase transitions between the MI and SF phases, as the coherent hopping amplitude is increased from zero. This is again surprising since it demonstrates that increasing the hopping amplitude, which in isolation favours the itinerancy of the bosons, can instead favour their localization under certain circumstances. In this work we analyse this unusual phenomenon of the 1D BHM directly in the thermodynamic limit by utilizing the infinite time-evolving block decimation (iTEBD) algorithm~\cite{Vidal07,Vidal04,Orus08} to variationally minimize the infinite matrix product state (MPS) ansatz~\cite{Rommer97}.

A highly unique feature of this family of states, heavily exploited here, is that it is parameterized by a matrix size $\chi$ which directly restricts the half-chain entanglement permitted in the state~\cite{Vidal07,Orus08}. This allows us to determine to what extent entanglement, which signals the presence of quantum correlations and fluctuations in the ground state, is essential for re-entrance to emerge. Further to this infinite MPS enable the application of the quantum information inspired finite-entanglement scaling~\cite{Tagliacozzo08} to study the KT transition at the MI lobe tip. In doing so we obtain an estimate of its location, derived from the behaviour of half-chain entanglement entropy, which is in excellent agreement with previous studies utilizing order parameters, energy gaps or correlations. This work thus provides further confirmation of the general applicability of this novel scaling procedure to non-integrable models.

The structure of this paper is as follows. In \secr{BHM} we give a brief overview of the properties of the BHM in 1D. In \secr{reentrance_itebd} we begin by describing re-entrance in the BHM, followed by the essential features of the infinite MPS ansatz employed here and how signatures of the MI-SF transition are manifested in our calculation. The MI lobes are then reported as a function of finite-entanglement from which we analyse the value of $\chi$ in which re-entrance is first observed. In \secr{kt_point}, after giving an overview of the extensive literature that has previously estimated the KT point in the 1D BHM, we then proceed to apply finite-entanglement scaling to obtain a new and complementary estimate on its location. Finally we conclude in \secr{conclusion}.

\section{Bose-Hubbard model}
\label{BHM}
The BHM Hamiltonian for a 1D chain in the grand-canonical ensemble is (taking $\hbar = 1$)
\begin{eqnarray}
\hat{H} &=& -t\sum_j\left(\hat{b}^{\dagger}_j\hat{b}_{j+1} + \textrm{H.c.}\right)  - \mu\sum_j\hat{n}_j + \frac{U}{2}\sum_j\hat{n}_j(\hat{n}_j-1), \nonumber
\end{eqnarray}
where $\hat{b}_j$ is the bosonic annihilation operator and $\hat{n}_j = \hat{b}_j^\dagger\hat{b}_j$ is the number operator at site $j$, respectively. Within $\hat{H}$ the chemical potential is given by $\mu$, while the kinetic energy is described by the hopping amplitude $t>0$ between neighbouring sites. In the absence of interactions hopping leads to a tight-binding energy band $\epsilon_{\textrm{kin}}(k) = -2t\cos(ka)$ with quasi-momentum $k$ and lattice spacing $a$. The repulsive interaction is described by the zero-range on-site term with positive strength $U$ which increases the energy if more than one boson occupies a given site. Throughout this work we set the energy scale to $U$ and for convenience label the ratios as $t/U \rightarrow t$ and $\mu/U \rightarrow \mu$. For numerical calculations we use a maximum occupation number of $n_{\textrm{max}} = 4$ boson per site, which for the near unit-filled calculations presented here is entirely sufficient.

\begin{figure}
\begin{center}
\includegraphics[width=9cm]{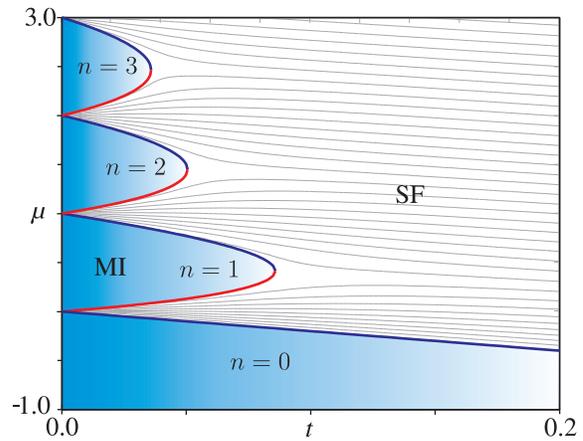}
\caption{(Color online) The mean-field phase $(t,\mu)$ diagram of the BHM in 1D depicting in the shaded regions the MI lobes with integer $n$ fillings~\cite{Fisher89,Sheshadri93,Rokhsar91,Krauth92}. These parabolic shaped lobes are surrounded by the SF phase. Some density contours in the SF phase are shown to illustrate their tendency to have a negative slope. The hole boundary is the lower side of the lobes while the particle boundary is the upper side. Note also that the MI lobes are particle-hole asymmetric.} \label{fig:mf_phase_diagram}
\end{center}
\end{figure}

Despite not being analytically solvable the form of the zero-temperature phase diagram of the BHM can be understood intuitively as follows~\cite{Fisher89,Sachdev01}. At $t = 0$ the ground state of the system is simply a product of on-site Fock states with no correlations. Every site is occupied by an integer number of bosons $n$ which minimizes the on-site energy $\epsilon_{\textrm{int}} = -\mu n + \half n(n-1)$. As a result within the interval of chemical potentials $n-1 < \mu < n$ the density is pinned at the integer $n$ and there is a finite interaction induced energy gap $\Delta(t=0) = \mu$ to the lowest-lying particle-hole excitation. This gapped state is a MI. As the hopping $t$ is turned on $\Delta(t)$ decreases, but its non-zero value is maintained for an extended region in the $(t,\mu)$ plane. Generally the energy gap $\Delta(t)$ is the distance in the $\mu$ direction between the particle (upper) and hole (lower) boundary and as it closes it gives rise to MI lobes. The familiar mean-field depiction~\cite{Fisher89,Sheshadri93,Rokhsar91,Krauth92} of the BHM phase diagram is shown in \fir{fig:mf_phase_diagram}.

Within the MI lobe correlations, such as $\av{\hat{b}^\dagger_0\hat{b}_x} \sim \exp(-x/\xi)$, are localized with a finite correlation length $\xi$ and, owing to the fixed density over a finite interval of $\mu$, the compressibility $\kappa = \partial \rho /\partial \mu$ vanishes. At some critical value of $t$, defining the boundary of the lobe, the kinetic energy overcomes the gap and there is a transition from the gapped, incompressible MI to a gapless, compressible SF phase surrounding the lobes. In 1D the SF possesses algebraically decaying correlations $\av{\hat{b}^\dagger_0\hat{b}_x} \sim x^{-K/2}$, with an exponent $K = \pi/\sqrt{\rho_s\kappa t}$, where $\rho_s$ is the superfluid density~\cite{Giamarchi04}. Since correlations decay asymptotically to zero the SF phase exhibits only quasi-long-range order and does not Bose condense. Nonetheless it is the existence of a non-vanishing $\rho_s$ which is the relevant criteria for superfluidity.

Only in the non-interacting limit $t \rightarrow \infty$ (or alternatively when $U=0$) does the system condense into the $k=0$ quasi-momentum state with every particle having an energy of $-2t$. Thus if the chemical potential $\mu < -2t$ then the system is empty (vacuum MI) since it costs energy to put a particle in, while for $\mu>-2t$ the number of particles goes to infinity because every additional particle reduces the total energy of the system. This tendency indicates that with non-zero interactions $U$, and $\mu$ fixed, increasing $t \rightarrow \infty$ will eventually drive the density to infinity~\cite{Elstner99b}. As such density contours in the $(t,\mu)$ plane will have a negative slope once $t$ is large enough, as illustrated in \fir{fig:mf_phase_diagram}.  

In order that the compressibility $\kappa$ is always positive the integer density contours of the SF phase must meet the tips of the corresponding MI lobes, implying that for the transitions across the tip the density remains commensurate. Correspondingly, the phase transition occurring at any other point on the lobe boundary, where a commensurately filled MI changes to an incommensurate filled SF, belongs to a different universality class from that at the tip. The scaling theory developed by Fisher {\em et al}.~\cite{Fisher89} showed that the tip is in the universality class of the $(d+1)$-dimensional XY model, whereas the generic transition is described by mean-field critical exponents in any dimension. In 1D this predicts that the lobe tips terminate with a KT transition point for which the gap closes asymptotically according to $\Delta(t) =A\exp(-B/\sqrt{t_{\rm KT} - t})$, where $t_{\rm KT}$ is the critical value of hopping, and with $A$ and $B$ being non-universal constants. Thus, the constant density KT transition is driven by phase fluctuations, while in contrast the generic transitions everywhere else on the lobe boundary are driven by density fluctuations.

\section{Re-entrance and entanglement}
\label{reentrance_itebd}

\begin{figure}
\begin{center}
\includegraphics[width=9cm]{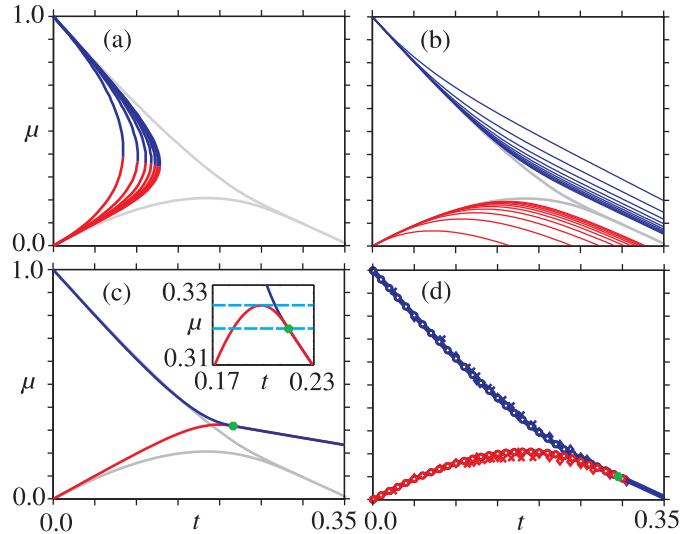}
\caption{(Color online) The phase diagram of the BHM predicted by a selection of different approximations. For (a), (b) and (c) the MI lobe determined from 12th order SCPE in Ref.~\onlinecite{Elstner99b} is also shown as the dotted line for comparison. In (a) the sequence of MI lobes are shown that were computed using mean-field decoupling~\cite{Fisher89,Sheshadri93,Rokhsar91,Krauth92} on the hopping between adjacent 1D $L$ site blocks varying in size from the conventional $L=1$ up to $L=8$ site clusters~\cite{Buonsante05}. In (b) the sequence of particle and hole boundaries computed from small finite-sized systems~\cite{Buonsante07}, with periodic boundary conditions, for $N=2$ to $N=11$ sites are shown. In (c) the real-space RG scheme described in \cite{Singh92} is used to determine the MI lobes~\cite{Pino12}, along with $\bullet$ marking the predicted KT point. The inset displays a zoomed in region of the phase diagram around the MI tip with horizontal dashed lines highlighting the marginal re-entrance. In (d) the DMRG results from \cite{Kuhner00} are shown as $\circ$, along with two different sets of QMC data from \cite{Kashurnikov96a} as $\times$ and \cite{Batrouni92} as $\Diamond$. The solid lines delininate the region found to have an unit density and the $\bullet$ marks the KT point found from the DMRG calculation in \cite{Kuhner00}.} \label{fig:phase_diagrams}
\end{center}
\end{figure}

In 1D the MI lobe have a pointed triangular shape strikingly different to the rounded mean-field lobe shown in \fir{fig:mf_phase_diagram}. Moreover the tips of MI lobes in 1D bend downwards giving them a signature ``claw'' shape and reflecting the presence of re-entrance not seen at all in higher dimensions. Crucial to both these features is the slowly closing gap at the tip's KT transition. This causes the tips location to be elongated to a much larger value of $t$ than mean-field predicts. That re-entrance occurs is then a combination of this property with the kinetic energy driven tendency for all density contours to acquire a negative slope with increasing $t$. For the unit-filled contour this effect is already manifested before the gap closes causing the lobe to follow this downward trend making the hole boundary concave.

Re-entrant behaviour of the MI lobes is a subtle feature which is often not captured accurately by commonly used approximations. For example in \fir{fig:phase_diagrams}(a) we show that neither a single-site decoupled mean-field theory~\cite{Fisher89,Sheshadri93,Rokhsar91,Krauth92}, nor its generalization to finite-sized 1D clusters~\cite{Buonsante05}, display any signs of re-entrants. However, some indications are visible in exact numerics for small finite-sized systems with periodic boundaries~\cite{Buonsante07}, as depicted in \fir{fig:phase_diagrams}(b) where the hole boundary is concave for all system sizes. We also show in \fir{fig:phase_diagrams}(c) that conventional real-space renormalization group applied to the 1D BHM does in fact predict a very marginal occurrence of re-entrance~\cite{Pino12}, a feature missed by the original work~\cite{Singh92}. Re-entrant behaviour was first convincingly demonstrated by an exhaustive 12th order strong coupling expansion (SCE) analysis \cite{Elstner99a}. For comparison their result is also plotted in \fir{fig:phase_diagrams}(a)-(c), illustrating how the finite-sized mean-field cluster in \fir{fig:phase_diagrams}(a) converge as a successive under-estimation, while the unclosed lobes of the finite-sized periodic systems in \fir{fig:phase_diagrams}(b) converge as a successive over-estimation. The presence of re-entrance was later resoundingly confirmed by several density matrix renormalization (DMRG)~\cite{Kuhner98,Kuhner00} and quantum Monte-Carlo (QMC) calculations~\cite{Batrouni92,Kashurnikov96a} whose original results are replotted in \fir{fig:phase_diagrams}(d). While the presence of re-entrance is well established here we give additional insight into its origin by exploiting unique characteristics of an infinite MPS approach.

\subsection{Finite-entanglement infinite MPS}
The matrix product state ansatz parameterizes the coefficients of a state $\ket{\Psi}$ of an $N$-site 1D lattice of $d$-dimensional quantum systems as a product of matrices. Specifically, an MPS  has a canonical form~\cite{Vidal07,Vidal04,Orus08}
\begin{equation}
\ket{\Psi} = \sum_{s_1\dots s_N} \tr\left[A_1(s_1)\Sigma_1A_2(s_2)\cdots A_N(s_N)\right]\ket{s_1\cdots s_N}, \nonumber
\end{equation}
where $s_i$ labels a basis for the local degree of freedom (for example boson number states) of site $i$, $A_i(s_i)$ are matrices associated to each site $i$, indexed by $s_i$, each with a fixed finite size $\chi$, and $\Sigma_i$ are positive diagonal matrices. Here we work directly in the thermodynamic limit $N \rightarrow \infty$, avoiding logarithmic finite size corrections seen in some earlier studies, and assume translational invariance. This yields a class of many-body states highly convenient for studying quantum phase transitions, called infinite MPS~\cite{Rommer97}, that are compactly described by just $O(d\chi^2)$ complex parameters contained in the site independent matrices $A(s_i)$ and $\Sigma$.

In principle to represent exactly an arbitrary state of an infinite lattice system requires $\chi \rightarrow \infty$. Instead the fixed matrix size $\chi$ within the infinite MPS ansatz is formally the maximum allowed rank of the Schmidt decomposition of $\ket{\Psi}$ when it is bipartitioned into two semi-infinite halves as~\cite{Vidal07,Orus08}
\begin{equation}
\ket{\Psi} = \sum_{\alpha=1}^\chi \lambda_\alpha\kets{\Phi^{[\triangleleft]}_\alpha}\kets{\Phi^{[\triangleright]}_\alpha}, \label{infinite_schmidt}
\end{equation}
where $\lambda_\alpha$ are real Schmidt coefficients, while $\kets{\Phi^{[\triangleleft]}_\alpha}$ and $\kets{\Phi^{[\triangleright]}_\alpha}$ are the orthonormal set of Schmidt states for the left and right half-chains. When an infinite MPS is in its canonical form the matrix elements of $\Sigma$ are the Schmidt coefficients $\lambda_\alpha$ and their importance is that they expose the entanglement between the two semi-infinite half chains as quantified by the von-Neumann entropy 
\begin{eqnarray}
S_{\frac{1}{2}} &=& -\sum_{\alpha=1}^\chi\lambda^2_\alpha\log_2(\lambda^2_\alpha). \label{half_vonneumann}
\end{eqnarray}
Thus, the maximum Schmidt rank $\chi$ permitted has a significant physical meaning as setting an upper limit $S_{\frac{1}{2}} = \log_2(\chi)$ to entanglement that can exist between two halves of the system within the description. The infinite MPS ansatz for $\chi>1$ therefore provides a powerful framework to go beyond mean-field theory by including non-trivial quantum correlations.  

Two powerful and highly efficient methods exist for computing directly an infinite MPS approximation to the ground state of a nearest-neighbour interacting 1D Hamiltonian like the BHM, namely the iTEBD~\cite{Vidal07,Orus08} and iDMRG algorithms~\cite{White92,McCulloch08}. While they operate on the same underlying ansatz~\footnote{Technically, both iTEBD and iDMRG operate on a infinite MPS ansatz which is two-site translationally invariant but will converge very close to a fully translationally invariant solution.} the main difference between them is that iTEBD is a power method of finding the ground state exploiting imaginary time-evolution, whereas iDMRG is a very efficient eigensolver method based on the diagonalization of an effective local Hamiltonian. For determining the most accurate infinite MPS approximation to the ground state the aim would be to use the largest accessible $\chi$, which is typically on the order of several 100's to 1000's. In this case the two algorithms tend to the same fixed point, but the iDMRG algorithm converges considerably faster~\cite{McCulloch08}. 

The aim of our work here is to instead examine systematically how the phase diagram of the BHM changes with increasing $\chi$, starting from its trivial $\chi = 1$ product state limit, and provide a rather unique perspective on the role of entanglement on its properties. As we shall describe shortly, the most profound changes to the structure of the MI lobe in fact occur at very small values of $\chi < 20$ where either algorithm converges fast. However, in the small $\chi$ regime the iDMRG algorithm is known to introduce a small perturbation, due to non-negligible truncation, causing it to produce a sub-optimal variational infinite MPS approximation~\cite{Dukelsky98,McCulloch08}. For this reason we instead employ the iTEBD algorithm whose imaginary time-evolution approach is highly robust allowing near optimal infinite MPS to be found irrespective of $\chi$. This ensures our results are insensitive to the specifics of the algorithm and instead reflect the underlying physics captured by an infinite MPS.

\subsection{Signatures of criticality at finite entanglement}
Despite operating in the thermodynamic limit any infinite MPS calculation with a finite $\chi$ can never display a genuine critical point due to its inability to produce the corresponding divergence in the correlation length~\cite{Rommer97}. Instead the finite-entanglement approach yields a pseudo-critical point whose location depends on $\chi$ but nevertheless provides useful information about the real transition point~\cite{Tagliacozzo08}. In this way we can characterize the pseudo transition at finite $\chi$ by computing a variety of both local and global quantities which display an anomaly of some form. In \fir{fig:qpt_anomalies}(a) the behaviour of some common quantities are shown as a function of $t$ along a line with $\mu = 0.60$ for a $\chi=3$ calculation. This includes local quantities, such as the density $\av{\hat{n}_j}$, its variance $\Delta(\hat{n}_j) = \av{\hat{n}^2_j} - \av{\hat{n}_j}^2$, the order parameter $\av{\hat{b}_j^\dagger}$, as well as a nearest-neighbour correlation $\av{\hat{b}^\dagger_j\hat{b}_{j+1}}_c = \av{\hat{b}^\dagger_j\hat{b}_{j+1}} - \av{\hat{b}_j^\dagger}\av{\hat{b}_{j+1}}$, and the global half-chain von-Neumann entropy $S_{\frac{1}{2}}$ of the state. All these quantities show an abrupt change at the same value of $t \approx 0.105$, whether it be the increase in $\av{\hat{n}_j}$ from unit filling expected for a generic MI-SF transition, or a corresponding local maxima in $S_{\frac{1}{2}}$. These changes ultimately reflect an abrupt change in the  nature of the ground state described by minimizing the infinite MPS ansatz. Owing to the constant and predictable behaviour of the density $\av{\hat{n}_j}$ in the MI phase, for all $\chi$, we use this quantity to isolate the lobe boundary and find that a precision of $\delta n=10^{-4}$ is sufficiently accurate away from the tip.

For $\chi = 1$ the infinite MPS ansatz reduces to an unentangled product state so minimization is equivalent to performing a self-consistent mean-field calculation. The boundaries determined by our procedure reproduce the analytic result for this limit. Moving beyond mean-field theory with a $\chi > 1$ cannot be performed for arbitrary values of $\chi$. Indeed when using $\chi = 2$ the iTEBD algorithm fails to converge, and instead $\chi = 3$ is the next smallest value. The reason for this numerical issue is entirely physical. The iTEBD algorithm, while not explicitly particle number conserving, nonetheless consistently minimizes, for any $\chi$, to a particle number symmetric infinite MPS when describing any ground state in the MI phase. The symmetry is numerically stable due to the finite gap in this phase. As a result of this symmetry the Schmidt coefficients $\lambda_\alpha$ for MI ground states posses a specific degeneracy structure, illustrated for a $\chi = 13$ calculation in \fir{fig:qpt_anomalies}(b). This shows that the 2nd and 3rd Schmidt coefficients are degenerate and thus any truncation to $\chi = 2$ will be unstable. The values of $\chi=1,3,5,7,8, \dots$ in which we truncate to are specifically chosen to respect this degeneracy structure.

In the gapless SF phase the infinite MPS minimized by the iTEBD algorithm does not preserve particle number symmetry, as illustrated in \fir{fig:qpt_anomalies}(a) by the order parameter $\av{\hat{b}_j}$ becoming non-zero. The finite-entanglement approach therefore inherits the mean-field theory feature of describing the MI to SF transition via symmetry breaking. As $\chi$ is increased the value of the order parameter $\av{\hat{b}_j}$ decreases suggesting that the exact non-symmetry breaking transition, where $\av{\hat{b}_j} = 0$ in both phases, is recovered in the $\chi \rightarrow \infty$ limit. However, unlike mean-field theory the Schmidt spectrum $\lambda_\alpha$ contained within the infinite MPS description is also an important signature of the transition. In \fir{fig:schmidt_spectrum} the Schmidt spectrum for a $\chi = 13$ calculation is displayed as a function of $t$. 

This `entanglement' spectrum has been proposed as a general method of detecting phase transitions which lie outside the Landau symmetry breaking paradigm. Based on the degeneracy structure of this spectrum a complete characterization of the topologically protected Haldane phase~\cite{Haldane88} of a $S=1$ spin chain~\cite{Pollmann10} and the Haldane insulator phase of the extended BHM~\cite{Deng11} have been demonstrated. In both cases it is argued that the transition to a topologically trivial phase is detected by the collapse of this structure. The utility of this entanglement spectrum approach is further confirmed here in \fir{fig:schmidt_spectrum} for the conventional MI-SF transition in the BHM. A splitting of Schmidt spectrum degeneracies in the MI phase, accompany the breaking of the global $U(1)$ particle number symmetry, is seen as the SF phase is entered around $t \approx 0.15$. Crucially \fir{fig:schmidt_spectrum} shows that a $\chi = 13$ calculation is already sufficiently entangled for the infinite MPS ansatz to display re-entrant behaviour. The reappearance of degeneracy in the Schmidt spectrum around $t \approx 0.20$ signals that the MI phase has been re-entered. We will now examine the shape of the MI lobe and the re-entrance phenomena as a function of $\chi$ in more detail.

\begin{figure}
\begin{center}
\includegraphics[width=9cm]{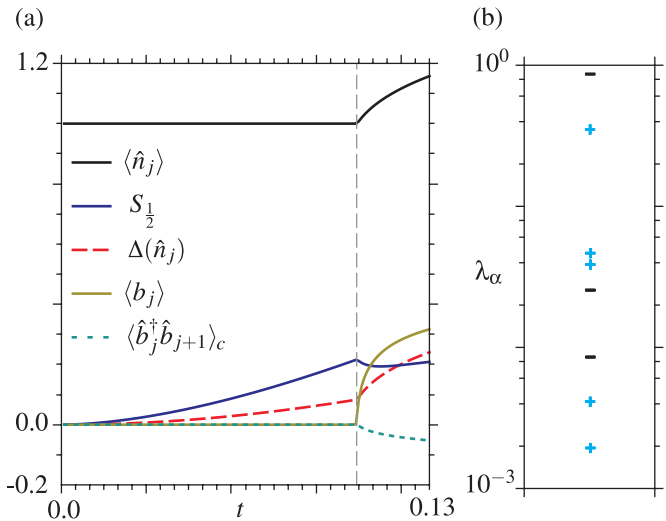}
\caption{(Color online) (a) Plotted here are for a $\chi = 3$ calculation are the on-site properties density $\av{\hat{n}_j}$, its variance $\Delta(\hat{n}_j)$, the order parameter $\av{\hat{b}_j^\dagger}$, as well as the nearest-neighbour correlation $\av{\hat{b}^\dagger_j\hat{b}_{j+1}}_c$, and the half-chain von-Neumann entropy $S_{\frac{1}{2}}$, as a function of $t$ with $\mu=0.60$. The order parameter and nearest-neighbour correlation are real since time-reversal symmetry ensures the ground state itself is real. All these quantities display anomalous kinks at the same pseudo-transition point (denoted by the vertical dashed line) and could in principle be used to detect the transition. (b) The Schmidt spectrum $\lambda_\alpha$ is displayed for a $\chi=13$ calculation inside the MI lobe at $\mu = 0.20$ and $t=0.22$. A non-degenerate Schmidt coefficient is represented by a `$-$', while a doubly degenerate one is given by `$+$'. This pattern of degeneracies is a consequence of $U(1)$ particle number symmetry being preserved by the infinite MPS ansatz when describing a MI and is found all over the lobe.} \label{fig:qpt_anomalies}
\end{center}
\end{figure}

\begin{figure}
\begin{center}
\includegraphics[width=9cm]{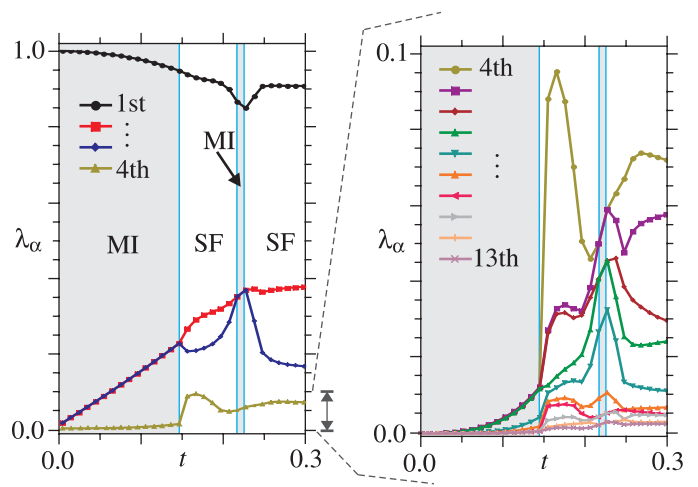}
\caption{(Color online) The Schmidt spectrum $\lambda_\alpha$ computed with $\chi = 13$ as a function of $t$ for constant $\mu=0.20$. In the left panel the four largest Schmidt coefficients are displayed. For $t \rightarrow 0$, deep in the MI regime, only one significant Schmidt coefficient $\lambda_1 \approx 1$ remains reflecting the unentangled unit-filled Fock state $\ket{\cdots 111 \cdots}$. As $t$ increases the infinite MPS description of the MI state becomes more complex with non-zero entanglement and a Schmidt spectrum possessing a degeneracy structure described in \fir{fig:qpt_anomalies}(b). When $t \approx 0.15$ the SF phase is entered and this degeneracy, visible here for $\lambda_2$ and $\lambda_3$, is lifted as the particle number symmetry is broken. This degeneracy splitting occurs right down the Schmidt spectrum, as shown in the zoomed in plot of the remaining Schmidt coefficients in the right panel. A $\chi = 13$ is sufficiently large that at this chosen $\mu$ re-entrance is reproduced by the finite-entanglement approximation. At $t \approx 0.20$ the MI phase is re-entered as signalled by the re-emergence of the degeneracy structure and particle number symmetry. In \fir{fig:qpt_anomalies}(b) the MI state shown lies inside this re-entrant region. For $t> 0.23$ this is finally broken once again as the SF phase is re-entered.} \label{fig:schmidt_spectrum}
\end{center}
\end{figure}

\subsection{Finite-entanglement lobes and re-entrance}
Using the abrupt changes in the density $\av{\hat{n}_j}$ we computed the lobe boundaries as a function of $\chi$ shown in \fir{fig:lobe_expansion}. This plot gives a systematic extrapolation of the MI lobe as a function of the maximum half-chain entanglement permitted. As $\chi$ is increased the lobe is seen to monotonically increase in size from the gross underestimate found in mean-field limit. This indicates that increased entanglement has a preferential effect on the description of the MI state, stabilising this phase for larger regions of $(t,\mu)$. Thus, in addition to symmetry breaking, much like the finite-cluster mean-field results shown in \fir{fig:phase_diagrams}(a), finite-entanglement calculations are found to predict lobes that are an underestimate of the exact one. Yet, quite remarkably the finite-entanglement lobes are seen to rapidly converge, to the essentially exact DMRG result~\cite{Kuhner00}, for extremely modest values of $\chi \sim 21$. 

Indeed the first jump from the $\chi = 1$ mean-field theory parabolic lobe to $\chi = 3$ infinite MPS lobe already accounts for $\approx 95\%$ of the exact lobe size. This is despite the fact that a $\chi = 3$ infinite MPS approximation never exhibits a correlation length much above $\xi \approx 2$ lattice sites, and so is substantially smaller than the finite-cluster sizes used in the mean-field approach. Furthermore a $\chi = 3$ infinite MPS has far fewer variational parameters than the $L=8$ site mean-field decoupled calculation. This illustrates that the infinite MPS ansatz even with a small $\chi$ is very adept at describing the relevant degrees of freedom. We can attribute this dramatic increase in the size of the MI lobe seen for $\chi = 3$ due to its nascent ability to describe an elementary particle-hole excitation, such as the superposition
\begin{equation}
\ket{\Psi} = \alpha \ket{\cdots 1111 \cdots} + \beta \left(\ket{\cdots 1201 \cdots} + \ket{\cdots 1021 \cdots}\right), \nonumber
\end{equation}
while retaining particle number symmetry and translational invariance. For small $\chi$ the tips of the lobes are still rounded enough that their location can be determined easily. This reveals another interesting feature, shown in \fir{fig:lobe_expansion}, that for all $\chi > 1$ the lobe tips intersect the line $\mu(t)=-1.538t+0.559$ to very good approximation. Thus for the region around these tips this line has yet to saturate to the $-2$ slope expected as $t\rightarrow \infty$. That the lobe elongates along such a well defined trajectory will be very useful shortly for performing a finite-entanglement scaling approximation on the KT critical point. It also indicates that re-entrance will emerge once the tip has moved sufficiently far along this line. 

\begin{figure}
\begin{center}
\includegraphics[width=9cm]{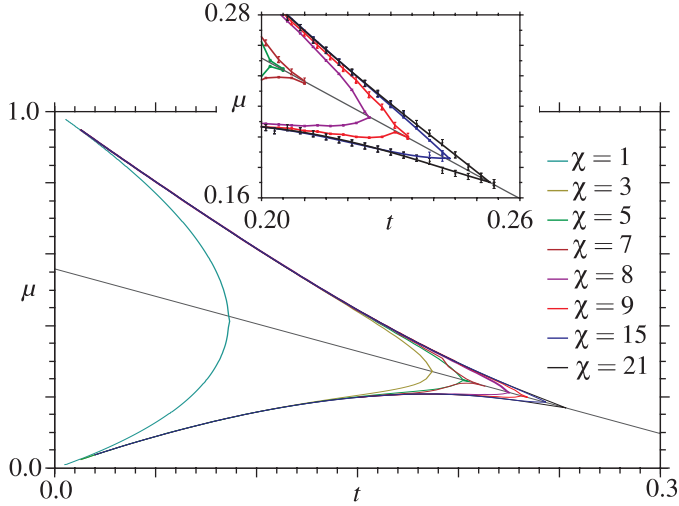}
\caption{(Color online) The boundaries of the MI lobe determined by iTEBD calculations for a sequence of Schmidt ranks $\chi$ ranging from a mean-field value of $\chi = 1$ to a sizeable finite entanglement value of $\chi = 21$. The inset shows a zoom in of the tip region for $\chi=5$ to $\chi = 21$. To good approximation for all $\chi>1$ shown the tips of these lobes intersect the line $\mu(t)=-1.538t+0.559$ displayed.} \label{fig:lobe_expansion}
\end{center}
\end{figure}

To determine the minimum value of $\chi$ required for re-entrance to appear we consider an enlarged plot of the tip region for the marginal cases of $\chi=7,8$ and $9$ in \fir{fig:re-entrance}. A clear distinction is seen between $\chi = 7$ and $\chi \geq 8$, as quantified by the inset of \fir{fig:re-entrance} which plots the derivative of the hole boundary. While $\chi = 8$ and $\chi = 9$ show a sizeable region of $t$ with a negative slope and concavity, $\chi = 7$ displays only a very small precursor near the tip. Thus, $\chi = 8$ is a measure of the minimum amount of entanglement needed for genuine re-entrance to manifest in the resulting MI lobe. That it is established at such a small value of $\chi$ is reminiscent of its similar emergence in SCE approaches where re-entrance is not seen for low orders, such as in an early 3rd order study~\cite{Freericks96}, and only appeared once a 10th order or higher expansions were performed~\cite{Elstner99b}. Here we can relate its emergence to the entanglement by examining the Schmidt spectrum. As shown in \fir{fig:qpt_anomalies} the 8th Schmidt coefficient in the MI phase is the first non-degenerate coefficient after the 3 degenerate pairs that follow the first Schmidt coefficient. The improvement in the description of the MI state induced by the corresponding Schmidt state for this coefficient therefore appears to be pivotal in establishing re-entrance. 

\begin{figure}
\begin{center}
\includegraphics[width=9cm]{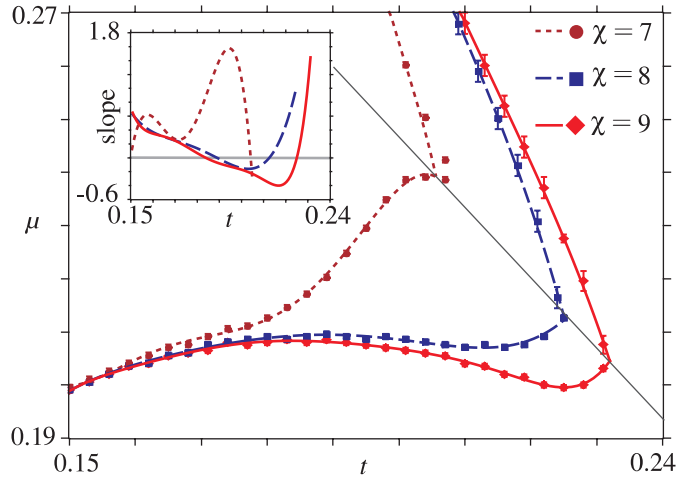}
\caption{(Color online) A zoomed in plot of the MI lobe boundaries for the cases $\chi=7,8$ and $9$ are shown. The hole boundary is fitted with a 4th order polynomial whose derivative, shown in the inset, quantifies the emergence of re-entrance. Both $\chi =8$ and $9$ show a significant finite region of $t$ where the slope is negative, indicating re-entrance, while $\chi = 7$ shows only a precursor to this with a very small negative slope at the cusp of the tip itself. The line $\mu(t)$ intersecting the tips is also shown} \label{fig:re-entrance}
\end{center}
\end{figure}

The shape of the MI lobe, and thus the appearance of re-entrance, is embodied by the underlying competition between the kinetic and interaction energies contained in the MI and SF states. To provide some physical insight into the influence of increasing $\chi$ we first rewrite the energy density $\epsilon$ of the BHM as
\begin{eqnarray}
\epsilon &=& -t\left(\av{\hat{b}^\dagger_j\hat{b}_{j+1}}_c  + \av{\hat{b}_j^\dagger}^2 + \textrm{c.c}\right) -\mu \av{n_j} \nonumber \\
&& + \half \Delta(\hat{n}_j) + \half \av{\hat{n}_j}(\av{\hat{n}_j} - 1), \nonumber
\end{eqnarray}
after exploiting translational invariance. In this form we expose the two contributions to kinetic energy, namely via symmetry breaking with a non-zero order parameter $\av{\hat{b}_j}$ and via inter-site correlations $\av{\hat{b}^\dagger_j\hat{b}_{j+1}}_c$. When either of these contributions are non-zero they induce non-zero on-site number fluctuations $\Delta(\hat{n}_j)$ contributing to the interaction energy. In the SF phase all these contributions will typically be non-zero. In general the correlation $\av{\hat{b}^\dagger_j\hat{b}_{j+1}}_c \neq 0$ only if the ground state is entangled. For the MI phase, however, particle number symmetry is highly restrictive prohibiting $\av{\hat{b}_j} \neq 0$ so any on-site number fluctuations $\Delta(\hat{n}_j) \neq 0$ also only appear if the ground state is entangled. Since the density is pinned to $\av{\hat{n}_j} =1$ these correlation and fluctuation contributions entirely account for the MI state energy density. For example, this means that at $\chi = 1$ the MI state simply has $\epsilon = -\mu$ independent of $t$, while in contrast the SF phase has the symmetry breaking mechanism available for it to accommodate increasing kinetic energy. This inevitably favours the gapless SF phase causing the underestimation of the MI lobe seen in mean-field theory. 

Once $\chi > 1$ the energetics of the MI state become much less trivial. To illustrate this in \fir{fig:energy_contributions} we plot the main contributions to $\epsilon$ as a function of $t$ for a $\chi=3,7$ and $\chi = 9$ infinite MPS. A $\mu = 0.204$ is chosen so that it cuts through the re-entrance which is present for $\chi=9$. It is apparent from \fir{fig:energy_contributions} that the description of on-site number fluctuations are not significantly different between the three approximations and grows monotonically with $t$ irrespective of the phase. Yet the emergence of a re-entrant MI state above $t > 0.20$ for $\chi = 9$ coincides entirely with a complete suppression of $\av{\hat{b}_j}$ along with a dramatic elevation in the correlation $\av{\hat{b}^\dagger_j\hat{b}_{j+1}}_c$ not seen for the lower $\chi$ states. Indeed for $\chi = 3$ the correlation contribution saturates for $t>0.10$. The importance of kinetic energy for re-entrance was implied earlier from its tendency to induce negative slope density contours. Here the energetics strongly suggests that it is the improved description of kinetic energy with increasing $\chi$ that is key to the emergence of re-entrance. 

We can understand the threshold in $\chi$ by noting that the nature of a ground state in general arises from the condition of having extremal local properties, like for the quantities contained in $\epsilon$, while simultaneously satisfying the global symmetries of the system. A classic example of this is a spin-$\half$ antiferromagnetic chain where in isolation each exchange interaction is minimized by a singlet state, however, frustration means that a spin cannot be in a singlet with both its neighbours. Instead, to recover translational invariance, the ground state becomes a complicated superposition of all singlet coverings creating quasi-long range order and requiring an infinite MPS description with $\chi > 1$~\cite{Verstraete08}. For the 1D BHM the re-entrant threshold for $\chi$ thus relates to the ability of the infinite MPS to describe the intricate correlations of a MI state possessing significant kinetic energy, while simultaneously satisfying both the translational and particle number symmetry. 
 
\begin{figure}
\begin{center}
\includegraphics[width=9cm]{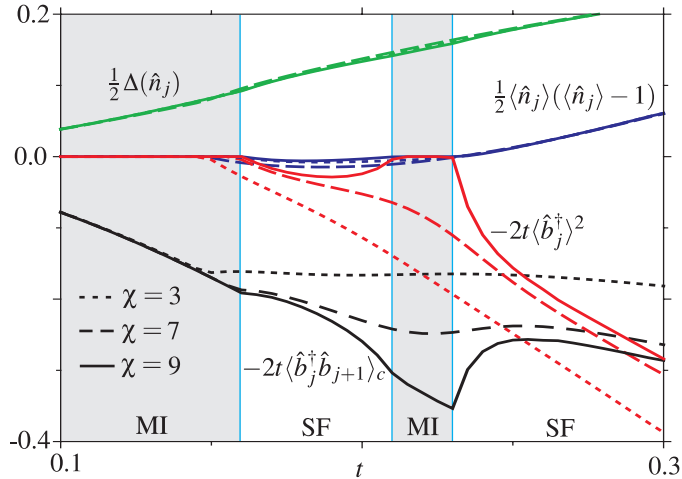}
\caption{(Color online) For $\chi=3,7$ and $9$ the fluctuation and correlation contributions to the ground state energy density $\epsilon$ are plotted as a function of $t$ for $\mu = 0.204$. Specifically the contributions are the half the on-site number variance $\half\Delta(\hat{n}_j)$, the uncorrelated doubly occupancy $\half\av{\hat{n}_j}(\av{\hat{n}_j}-1)$, the order parameter squared $-2t\av{\hat{b}_j^\dagger}^2$ and the nearest-neighbour correlation $-2t\av{\hat{b}^\dagger_j\hat{b}_{j+1}}_c$. The phase boundaries depicted by the shading are those determined by $\chi = 9$, which displays re-entrance. While the  on-site interaction terms display only marginal changes with $\chi$, the kinetic energy contributions change substantially. In the re-entrant region a $\chi = 9$ is sufficient to describe increased kinetic energy via the correlation $\av{\hat{b}^\dagger_j\hat{b}_{j+1}}_c$ while suppressing entirely the symmetry breaking contribution $\av{\hat{b}_j^\dagger}^2$.} \label{fig:energy_contributions}
\end{center}
\end{figure}

\section{The Kosterlizt-Thouless transition}
\label{kt_point}

\subsection{Overview of KT critical point calculations}
Due to its infinite order and associated very slowly closing energy gap $\Delta(t)$, the essential singularity of the KT transition is notoriously hard to compute numerically. For this reason numerous studies have utilised a variety of different methods to tackle its characterisation. To put our estimate, described shortly, into context we give here a brief overview of these extensive findings.

By far the simplest approach is a site-decoupled mean-field theory~\cite{Fisher89,Sheshadri93,Rokhsar91,Krauth92} which gives $t_{\rm KT} = 0.086$ seen already in \fir{fig:mf_phase_diagram}, a result which becomes exact in infinite-dimensions. By considering a truncated BHM, where no triple or higher occupancies are allowed, a real-space renormalization group approach was applied in Ref.~\cite{Singh92} giving $t_{\rm KT} = 0.215$, as seen in \fir{fig:phase_diagrams}(c). Early quantum Monte Carlo (QMC) calculations~\cite{Batrouni90,Batrouni92} also estimated from the closing energy gap essentially the same value $t_{\rm KT} = 0.215 \pm 0.02$. However, an analytical approach~\cite{Krauth91}, based on the Bethe ansatz approximating the full BHM, but conjectured to correspond to the exact solution of the truncated model, yielded a much higher estimate $t_{\rm KT} = 0.289$. Furthermore, other exact diagonalization scaling studies of the full BHM, like that in \fir{fig:phase_diagrams}(b), also found $t_{\rm KT} = 0.275 \pm 0.005$ using the energy gap $\Delta(t)$~\cite{Park04}, $t_{\rm KT} = 0.283 \pm 0.005$ from the SF stiffness $\rho_s$~\cite{Elesin94}, and $t_{\rm KT}=0.257 \pm 0.001$ from the derivative of the ground state fidelity~\cite{Buonsante07}. 

Later calculations corroborated this larger value of $t_{\rm KT}$ with a more recent QMC calculation~\cite{Kashurnikov96a} giving $t_{\rm KT} = 0.300 \pm 0.005$, very closely followed by an exact diagonalization study combined with renormalization group~\cite{Kashurnikov96b} giving $t_{\rm KT} = 0.304 \pm 0.002$. A quite different approach from these has been pursued using SCE on the full BHM. Early 3rd order expansions~\cite{Freericks96} gave a bare estimate of $t_{\rm KT} = 0.215$ which was then modified to $t_{\rm KT} = 0.265$ once a careful extrapolation was made taking account of the KT nature of the tip~\cite{Elstner99a}. Subsequent related work dramatically enhanced the SCE to 12th order to give $t_{\rm KT} = 0.26 \pm 0.01$~\cite{Elstner99b}. 

Some of the most accurate treatments of this problem have utilised the DMRG method~\cite{White92}. The first such study used the infinite-size DMRG algorithm with periodic boundary conditions to compute the energy gap $\Delta(t)$ allowing an extrapolation to find $t_{\rm KT} = 0.298 \pm 0.01$~\cite{Pai96}. Owing to the difficulties associated with $\Delta(t)$ a second study instead computed, with the same algorithm, the behaviour of long-range correlations $\av{\hat{b}^\dagger_0\hat{b}_x}$ across the KT transition. From Luttinger liquid theory the asymptotic exponent $K$ of these correlations is known to be exactly $K=\half$ at the critical point and this enabled a different estimate to be found as $t_{\rm KT} = 0.277\pm 0.01$~\cite{Kuhner98}. Another study, using instead the finite-size DMRG algorithm with periodic boundary conditions, found a slightly lower estimate of $t_{\rm KT} = 0.26$ closer to those determined with SCE~\cite{Rapsch99}. One of the most exhaustive DMRG calculations, following these previous works, used finite-size algorithm with open boundary conditions~\cite{Kuhner00}, where DMRG is known to be most accurate. It again examined the decay of the $\av{\hat{b}^\dagger_0\hat{b}_x}$ correlation for large finite-sized systems giving an estimate of $t_{\rm KT} = 0.297\pm 0.01$ higher than most earlier values. 

That $t_{\rm KT}$ has been underestimated in the very early works has been revealed by numerous recent calculations. One of these exploits the iTEBD algorithm used here to estimate~\cite{Zakrzewski08} the KT critical point as $t_{\rm KT} = 0.2975 \pm 0.0005$, once again from the decay of the $\av{\hat{b}^\dagger_0\hat{b}_x}$ correlation. Other recent finite-sized DMRG calculations have used finite-size scaling to estimate $t_{\rm KT} = 0.2980 \pm 0.0005$ from the von-Neumann entropy~\cite{Lauchli08}, $t_{\rm KT} = 0.3050 \pm 0.001$ from the density-density $\av{\hat{n}_0\hat{n}_x}$ correlation~\cite{Ejima11}, $t_{\rm KT} = 0.3030 \pm 0.009$ from the energy gap~\cite{Roux08}, $t_{\rm KT} = 0.3190 \pm 0.001$ from the winding number excitation gap~\cite{Danshita11} and $t_{\rm KT} = 0.2989 \pm 0.0002$ from bipartite density fluctuations~\cite{Rachel12}. As is apparent, these works have produced a spread of estimates for the critical hopping $t_{\rm KT}$ with non-overlapping error bars, indicating that not all the dominant uncertainties in these results have been accurately accounted for. Nonetheless, these most recent and extensive studies have established a consensus, by examining a variety of different quantities, that $t_{\rm KT} \approx 0.3$. 

Our approach here has similarities to these recent studies in that it uses an MPS ansatz and a similar minimization algorithm, but also differs substantially in that we work directly in the thermodynamic limit and employ finite-entanglement scaling~\cite{Tagliacozzo08}, as opposed to finite-size scaling. As we shall now show this does not rely on the asymptotic behaviour of any specific correlation function or energy gap. Instead it is based on general scaling arguments about the half-chain entropy $S_{\frac{1}{2}}$ that apply irrespective of many of the microscopic details of the underlying model. 

\subsection{Entanglement scaling of the KT point}
As we saw earlier, for a fixed $\chi$ an infinite MPS approximation produces pseudo-critical points $t_c(\chi)$, whose location depends on $\chi$ and is signalled by the singular behaviour of certain physical quantities. Due to this the application of scaling analysis with $\chi$ in the region near the transition was proposed~\cite{Tagliacozzo08} as a means of extracting information about the nature and location of actual critical phenomenon in the limit $\chi \rightarrow \infty$. So far this approach has been successfully applied to the transverse field Ising model and Heisenberg model~\cite{Tagliacozzo08}, as well as the transverse axial next-nearest-neighbor Ising model~ \cite{Nagy11}. Here we apply it for the first time to accurately determine the KT point of the unit-filled MI lobe.

To determine the pseudo-critical point at the tip of the lobes we assume that with increasing $\chi$ the trajectory of the tips towards the true KT point is given by the line $\mu(t)$ in the $(t,\mu)$ plane defined earlier. Previous calculations, shown in \fir{fig:lobe_expansion}, confirmed this to be a good approximation for $3 \leq \chi \leq 21$. In contrast to the generic transitions used to constrain the lobe structure, locating the pseudo-critical point at the tip, where $\av{\hat{n}_j}$ remains fixed at unity, is a more difficult task. For this reason we instead exploit the local maxima anomaly in $S_{\frac{1}{2}}$, shown earlier in \fir{fig:qpt_anomalies}(a) for a generic transition. In \fir{fig:entanglement_scaling}(a) the half-chain entropy $S_{\frac{1}{2}}$ is shown as a function of $t$ along the line $\mu(t)$ for a sequence of finite-entanglement calculations ranging from $\chi=7$ (bottom) to $\chi=70$ (top). The local maxima anomaly persists at the tip with a value increasing with $\chi$ suggestive of a divergence in $\chi \rightarrow \infty$ limit. From these maxima we extract the pseudo-critical hopping $t_c(\chi)$ for each $\chi$ providing the raw data necessary for finite-entanglement scaling. From this data alone it is possible to perform a crude scaling analysis, such as a simple power-law extrapolation, without recourse to prior knowledge of the nature of the transition, and obtain a reasonable initial estimate of its location. However, to proceed more accurately we require an appropriate scaling relation, taking into account the known KT nature of the transition, describing the asymptotic functional form of $t_c(\chi)$ .

\begin{figure}
\begin{center}
\includegraphics[width=9cm]{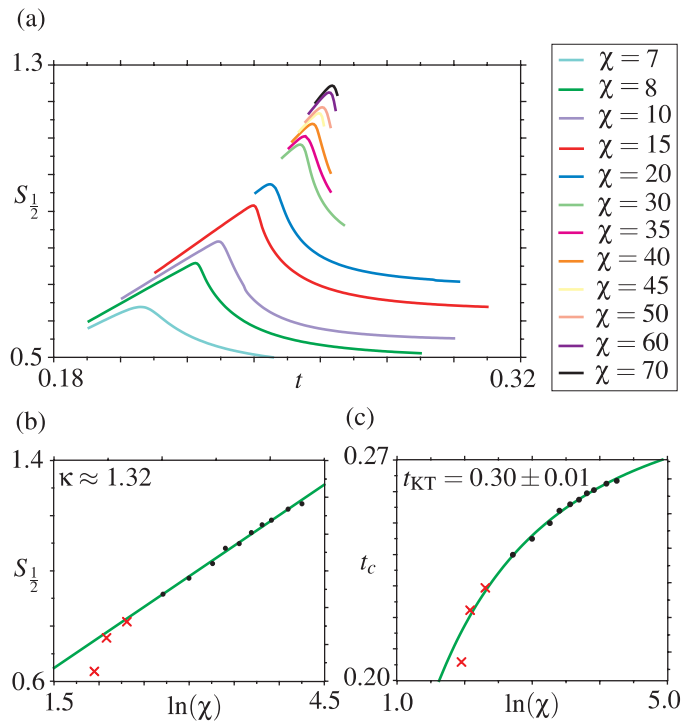}
\caption{(Color online) (a) The half-chain von Neumann entropy $S_{\frac{1}{2}}$ computed as a function of hopping $t$ along the line $\mu(t)$ depicted in \fir{fig:lobe_expansion} for a sequence of different values of $\chi$ ranging from $\chi=7$ (bottom curve) to $\chi = 70$ (top curve). The maximum value of $S_{\frac{1}{2}}$ increases as a function $\chi$ as more entanglement is permitted within the infinite MPS ansatz. (b) A plot of the value of $S_{\frac{1}{2}}$ at the pseudo KT point as a function of $\ln(\chi)$. The points are least-squared fitted by a line $A_{1}\ln(\chi)+A_{2}$ with $A_{1}=0.22$ and $A_{2}=0.32$. (c) The location of the pseudo-critical points $t_{c}(\chi)$ as a function of $\ln(\chi)$. The line fitting the points (with a chi-squared $= 12.5$ and $7$ degrees of freedom) is given in \eqr{eq:t_chi_kt}, with the free parameters found to be $B_{1}\approx -1.22$, $B_{2}\approx 1.93$ and $t_{\rm KT}=0.30\pm 0.01$. In both fittings for (b) and (c) the three smallest values of $\ln(\chi)$, corresponding to $\chi=7,8$ and $10$, have been excluded. Note how these points (denoted by $\times$) deviate considerably from the fit indicating that the asymptotic scaling regime is not satisfied for such small values of $\chi$.} \label{fig:entanglement_scaling}
\end{center}
\end{figure}

As observed earlier in \fir{fig:lobe_expansion} the MI lobe monotonically increases in size with increasing $\chi$. As such the pseudo-critical point $t_c(\chi)$ at the tip of the lobe for any finite $\chi$ will be an underestimation of the exact KT point and the exact phase of the system at $t_c(\chi)$ is MI with a finite correlation length $\xi_c(\chi)$. Following Ref.~\onlinecite{Tagliacozzo08} we assume that the relation between the correlation length $\xi_c$ and $\chi$ is of the form
\begin{equation}
\xi_c(\chi) \sim \chi^\kappa, \label{eq:xi_chi_scaling}
\end{equation}
where $\kappa$ is a real exponent. To test this relation we use the scaling of the half-chain entropy $S_{\frac{1}{2}}$ with the correlation length near a critical point found in conformal field theory~\cite{Calabrese04,Vidal03,Holzhey94} as
\begin{equation}
S_{\frac{1}{2}} \sim \frac{c}{6} \ln\left(\frac{\xi}{a}\right), 
\end{equation}
where $c$ is the corresponding central charge. Assuming that \eqr{eq:xi_chi_scaling} holds we expect the dependence of $S_{\frac{1}{2}}$ on $\chi$ at the pseudo-critical point to be
\begin{equation}
S_{\frac{1}{2}} \sim \frac{\kappa c}{6} \ln(\chi). \label{eq:s_chi_scaling}
\end{equation}
In \fir{fig:entanglement_scaling}(b) we plot the maxima of $S_{\frac{1}{2}}$ extracted from \fir{fig:entanglement_scaling}(a) and a fitted line $S_{\frac{1}{2}}=A_{1}\ln{\chi}+A_{2}$. The asymptotic nature of the scaling relations implies that it should only be fitted for sufficiently large $\chi$. Our data suggests that the scaling regime for finite-entanglement scaling can be attained by using pseudo-critical points for $\chi>10$. While this is a small $\chi$ physically its corresponds to including only the tips of MI lobes where re-entrance has already been firmly established, signifying the fundamental connection between these phenomenon. We find that the $\chi > 10$ points fit well to \eqr{eq:s_chi_scaling}, with a slope $A_{1}\approx 0.22$, indicating the validity of the scaling relation \eqr{eq:xi_chi_scaling}. Since the central charge for a KT transition is $c=1$ we then estimate that $\kappa \approx 1.32$ for the BHM. This value is in close agreement to that found in Ref.~\cite{Tagliacozzo08} for the KT critical point in the Heisenberg model. This coincidence further confirms that exponent $\kappa$ depends only on the nature of the critical point and not on the microscopic details of the underlying Hamiltonian. 

Having established the applicability of \eqr{eq:xi_chi_scaling} we now employ the known KT scaling relation~\cite{Kosterlitz73,LeBellac04} connecting the pseudo-critical point $t_c$ to the correlation length $\xi_c$
\begin{equation}
t_c(\xi_c) = \frac{C_ {1}}{\left(\ln{\xi_c}+C_{2}\right)^{2}}+t_{\rm KT}, \nonumber
\end{equation}
where $C_1$ and $C_2$ are real constants and $t_{\rm KT}$ is the location of the true KT point. Using \eqr{eq:xi_chi_scaling} again we then obtain the corresponding scaling of $t_c$ with
the entanglement as
\begin{equation}
t_{c}(\chi) = \frac{B_{1}}{\left(\ln{\chi}+B_{2}\right)^{2}}+t_{\rm KT},  \label{eq:t_chi_kt}
\end{equation}
with $B_{1}$ and $B_{2}$ real constants. In \fir{fig:entanglement_scaling}(c) we plot $t_c$ against $\ln(\chi)$ and the fitting to \eqr{eq:t_chi_kt}. This analysis yields an estimate
\begin{equation}
t_{\rm KT} = 0.30 \pm 0.01. \nonumber
\end{equation}
Our result shows excellent agreement to the large body of recent studies~\cite{Kuhner00,Zakrzewski08,Ejima11,Rachel12,Roux08,Lauchli08} reviewed earlier. While complementary to those calculations, the finite-entanglement scaling origin of this estimate is quite distinct from the fitting of a specific correlation function or energy gap performed in those previous studies. As such we have shown that this approach can also yield a precise critical point even in this most demanding case. 

The generous error bars in our estimate are derived from fitting \eqr{eq:t_chi_kt} to the data. This is the dominant contribution to the uncertainty in our estimate since the data itself for each $\chi$ was confirmed to be essentially exact by converging the imaginary time-evolution in iTEBD sufficiently well with a sequence of decreasing time-steps. Another much smaller source of error, not explicitly accounted for, arises from assuming that the tip lies on the line $\mu(t)$. Deviations from this line would result in the pseudo-critical point located being a generic transition in close proximity to the tip, rather than the tip transition itself, and therefore underestimating its value. To constrain this we confirmed that for the largest $\chi$ considered our calculations produced a $\av{\hat{n}_j}$ which remained unchanged, to within the precision $\delta n$, when crossing the transition identified. An obvious strategy to improve this finite-entanglement estimate would be to perform a more exhaustive and higher precision determination of both the $t$ and $\mu$ location of the pseudo-critical point in this region. Additionally the use of larger $\chi$ data would increase the data set size used in the fitting and also further ensure that the scaling-regime was entered. This could substantially reduce the uncertainty and potentially yield one of the most accurate determinations of $t_{\rm KT}$ free from logarithmic finite-sized corrections.

\section{conclusions}
\label{conclusion}
Utilizing the iTEBD algorithm we have performed a finite-entanglement analysis of the MI-SF transition in the BHM in 1D. The infinite MPS ansatz applied provides a unique extrapolation beyond mean-field theory in which a restriction on the entanglement between the two semi-infinite half chains is the defining characteristic. Crucially by operating directly in the thermodynamic limit this approach simplifies a scaling analysis approach since only the entanglement needs to be considered. The infinite MPS description enabled us to study the influence of entanglement on the uniquely 1D characteristics of the MI-SF transition such as re-entrance in the MI lobe shape and the location of the KT transition at its tip. We found that the minimum Schmidt rank in which re-entrance was manifested in the MI lobe was $\chi \geq 8$. The physical origins of this threshold was shown to be connected to the entanglement needed for an infinite MPS to be both particle-number symmetric and effectively capture intricate particle-hole excitations above the MI state carrying kinetic energy. We then focused on the tip of the lobe and performed a finite-entanglement scaling analysis of the infinite order KT critical point known to exist there. In this approach
we found that using a $\chi$ sufficiently large so that re-entrance was already present allowed the scaling regime to be reached to good approximation. By using the location of the pseudo-critical point up to a very moderate $\chi = 70$ we obtained a new estimate of the KT point as $t_{\rm KT}=0.30\pm 0.01$, in excellent agreement with the best earlier works based on DMRG. This illustrates how relatively low cost iTEBD calculations can not only provide qualitative insight into the lobe structure but, in combination with the appropriate scaling relations, also provides quantitatively accurate estimations of critical points via finite-entanglement scaling. While our work has highlighted tentative links between the KT nature of the tip transition in 1D and re-entrance, open questions still remain about how deep this connection is.  Future work employing an explicitly particle-number conserving iTEBD approach~\cite{McCulloch08}, thereby guaranteeing that the KT point is crossed numerically, might reveal further insight.

\acknowledgements
SRC and DJ thank the National Research Foundation and the Ministry of Education of Singapore for support. JP was supported by Ministerio de Ciencia e Innovaci\'{o}n Project No. FIS2009-13483-C02-02 and the Fundaci\'{o}n S\'{e}neca Project No. 11920/PI/09-j. MP and AMS acknowledge support from DGI Grant No. FIS2009-13483.

\pagebreak

\end{document}